# Multiscaling for Systems with a Broad Continuum of Characteristic Lengths and Times: Structural Transitions in Nanocomposites

S. Pankavich[a,b] and P. Ortoleva[a]

Center for Cell and Virus Theory[a]

Department of Chemistry

Indiana University

Bloomington, IN 47405

Department of Mathematics[b]

University of Texas at Arlington

Arlington, TX 76019

Contact: ortoleva@indiana.edu

**ABSTRACT**

The multiscale approach to $N$-body systems is generalized to address the broad continuum of long time and length scales associated with collective behaviors. A technique is developed based on the concept of an uncountable set of time variables and of order parameters (OPs) specifying major features of the system. We adopt this perspective as a natural extension of the commonly used discrete set of timescales and OPs which is practical when only a few, widely-separated scales exist. The existence of a gap in the spectrum of timescales for such a system (under quasi-equilibrium conditions) is used to introduce a continuous scaling and perform a multiscale analysis of the Liouville equation. A functional-differential Smoluchowski equation is derived for the stochastic dynamics of the continuum of Fourier component order parameters. A continuum of spatially non-local Langevin equations for the OPs is also derived. The theory is demonstrated via the analysis of structural transitions in a composite material, as occurs for viral capsids and molecular circuits.

Keywords: multiscale analysis, continuous scaling, Liouville equation, functional expansion, nanoparticles, nanomaterials, molecular circuits, self-assembly, structural transitions

## 1 INTRODUCTION

There are many systems composed of multi-atom, nanoscale components that can assemble into a composite structure. Examples include:

- nanoparticle arrays at solid or liquid interfaces
- networks of protein filaments (e.g. the cytoskeleton or the spindle for chromosome separation)
- composite three-dimensional materials made of macromolecules or nanoparticles
- clusters of cell surface receptors
- a viral capsid assembled from proteins organized into protomers, pentamers, or hexamers.

Due to the large number of atoms in each nanocomponent, the center-of-mass (CM) and orientation evolve much more slowly than the $10^{-14}$ second timescale of atomic vibration or collision. In addition to the atomistic high-frequency phenomena, the dynamics evolve on a wide range of long timescales associated with the collective modes of the multi-component system, including the coherent response of dimers, trimers, and larger clusters of components. These considerations suggest the existence of a gap between the short timescale ($10^{-14}$ second) and the broad range of long timescales of the coherent motion of single or multiple nanometer size components, the longest timescale being associated with the coordinated motion of many components simultaneously. In this way, such composite systems can display a wide spectrum of timescales starting with the microscopic time, followed by a gap, and then a broad quasi-continuum of long timescales. For this reason, traditional two or three timescale approaches cannot be applied.

It has long been known that an *N*-body system can display phenomena spanning a wide range of scales in both space and time. For example, hydrodynamic modes can exist over a broad range of scales. Phase transitions in viral capsids are believed to nucleate with the displacement and rotation of a single protein or protomer and, subsequently, progress across the capsid as a structural transition front. Similarly there is a broad range of vibrational modes of nanocomposite materials corresponding to the vibrations of clusters of nanocomponents. The traditional approach to multiscale systems is to introduce a timescale ratio $\varepsilon$ (<<1) and an associated sequence of time variables $t_n = \varepsilon^n t$, $n = 0, 1, \cdots$ that capture a set of processes, each with a discrete characteristic time and separated from nearby scales by a gap. However, this is not appropriate when there are so many very closely packed timescales as to form a quasi-continuum extending over many orders of magnitude. In a sense, one expects an approach based on a more general set of time variables, $t_n$ indexed not by the set of non-negative integers, but instead an uncountable set (e.g., positive numbers).

One approach to multiscale systems that seems to avoid discrete scaling is the projection operator method[7]. This method accounts for slow variables not included in a reduced description of interest, e.g., hydrodynamic modes in a host fluid interacting with a nanoparticle. Unfortunately, the memory kernels introduced by this method cannot readily be evaluated, although insights can be obtained by computing them approximately or fitting them to experimental data. The evaluation of these memory kernels requires Molecular Dynamics (MD) simulations on the order of the timescale of the kernels. If this is long, then the calculations are impractical. Instead, if this is short then the approach presented in the following sections will directly yield the reduced equations of interest; thus removing the need to use projection operators. Practical progress can be made within the projection operator framework when it is combined with mode-coupling theory. However, such an analysis does not take advantage of the gap in the spectrum of timescales often displayed by *N*-body systems. For example, a disturbance created by launching a nanoparticle in a host medium of small molecules could excite only those modes of the host with wavelengths greater than or equal to the size of the nanoparticle or with timescales greater than or equal to the size of the nanoparticle divided by its maximum velocity. The objective of the present study is to develop an analysis of *N*-body systems that captures the character of this gap/continuum scaling phenomena.

The multiple scale understanding of *N*-body systems dates back to Einstein, Langevin, Smoluchowski, Chandrasekhar and others[5, 6, 8, 10]. Deutch and Oppenheim[7] presented an approach to Brownian motion based on projection operators and a perturbation scheme in the host particle/nanoparticle mass ratio. This approach set the stage for a series of analyses of Brownian particle dynamics based on the Liouville equation leading to the derivation of Fokker-Planck (FP) equations. Shea and Oppenheim[9] derived FP and Langevin equations for a single structureless Brownian particle in a host medium using projection operators and perturbation techniques based on expansions in the mass ratio or macroscopic gradients in the host medium. The original approach of Ortoleva et al[11-16] is based on a formal discrete multiple space-time scaling approach integrated with a statistical argument derived from the host particle/Brownian particle size ratio that allows for a united asymptotic expansion to arrive at FP equations for single and multiple Brownian particles and intra-Brownian particle structural dynamics. When the slow hydrodynamics of the host medium are accounted for, it was shown formally that the treatment leads to a set of coupled FP equations, one for each host mode. Shea and Oppenheim[9] extended their work to the case of multiple nanoparticles, introducing a number of smallness parameters. In [11-16] a unified asymptotic expansion and a multiple time scale formulation was presented that captures transitional behaviors (e.g. when the timescale switches as the number of nanoparticles per volume increases so that the behavior becomes more like that of a solid porous matrix than a suspension of disconnected nanoparticles). The notion of space-warping was introduced to treat nanoparticle deformation and rotation[17]. Additionally, we have placed the space-warping method in a self-consistent framework for analyzing single and multiple nanocomponents with internal atomistic structure. This will be shown in the present study to facilitate the development of a continuum multiscaling theory[12-16,18-19].

In all the above multiscale treatments, only a discrete set of time scales, ordered by powers of a small parameter $\varepsilon$, was considered. The existence of a broad continuum of scales has not been addressed even though such behavior is prevalent in many systems, so that the development of some generalization of discrete scaling is required.

In this study we address systems for which there is a set of short timescale processes, and also a broad continuous range of long timescales. The discrete problem is briefly reviewed in Sect 2. In Sect. 3, the continuum scaling problem for a set of interacting nanocomponents is formulated, and a perturbation analysis is performed. In Sect. 4 we derive a spatially non-local Langevin equation that is equivalent to the functional Smoluchowski equation of Sect. 3 and discuss how it can be used to simulate systems with a broad spectrum of length and time scales. Conclusions are drawn in Sect. V.

## 2 ORDER PARAMETERS, EFFECTIVE MASS, AND DISCRETE SCALING

Consider a set of *M* identical nanocomponents, each containing many atoms. Let the mass of each component scale as $\varepsilon^{-1}$ times a typical atomic mass *m* for $\varepsilon << 1$. The components are immersed in a host fluid, the entire system constituting a total of *N* atoms. The objective of this section is to demonstrate the breakdown of classic multiscale techniques and the need for continuous scaling in such a system when *M* is large. We do so in the course of deriving an equation for $\Psi$, the reduced probability density for $\underline{R} = \left\{\vec{R}_1, \ldots, \vec{R}_M\right\}$, the set of CMs of the nanocomponents.

The reduced probability density $\Psi$ is defined via

$$\Psi\left(\underline{R}, t\right) = \int d^{6N}\Gamma^* \Delta\left(\underline{R} - \underline{R}^*\right)\rho, \tag{2.1}$$

where $\rho$ is the *N*-atom probability density, $\Delta$ is an *M*-fold product of Dirac delta functions, and $\Gamma$ describes the state of the *N*-atom system and $\Gamma^*$ is the *N*-atom state over which integration is taken; $\underline{R}^*$ is an expression for the $M$ CMs in terms of the *N*-atom configuration $\Gamma^*$. This latter, starred, notation will be used below to denote other quantities expressed in the integration variables. Here, $\rho$ satisfies the Liouville equation:

$$\frac{\partial \rho}{\partial t} = \mathcal{L}\rho; \qquad \mathcal{L} = -\sum_{i=1}^{N}\left(\frac{\vec{p}_i}{m_i} \bullet \frac{\partial}{\partial \vec{r}_i} + \vec{F}_i \bullet \frac{\partial}{\partial \vec{p}_i}\right) \tag{2.2}$$

where $m_i$, $\vec{p}_i$, $\vec{r}_i$, and $\vec{F}_i$ are the mass, momentum, position, and force on atom $i$.
Using (2.1),(2.2), and integration by parts, we have

$$\frac{\partial \Psi}{\partial t} = -\varepsilon \frac{\partial \underline{J}}{\partial \underline{R}} \tag{2.3}$$

$$\underline{J} = \frac{1}{m}\int d^{6N}\Gamma^* \Delta\left(\underline{R} - \underline{R}^*\right)\underline{P}^* \rho\left(\Gamma^*, t\right) \tag{2.4}$$

where $\underline{P} = \left\{\vec{P}_1, \ldots, \vec{P}_M\right\}$ is the set of total momenta of the *M* nanocomponents. We determine conditions under which the conservation law (2.2) is closed in $\Psi$, i.e., involves only the reduced distribution $\Psi$ and not the full *N*-atom distribution $\rho$.

Similar to methods developed earlier[12-16], $\rho$ is envisioned to depend on $\Gamma$ directly and through the *M* nanocomponent center-of-mass (CM) positions $\underline{R}$ , indirectly. For simplicity, we ignore the orientation of the nanocomponents in this exploratory study. With this ansatz on the dual dependence of $\rho$ on $\Gamma$ and the chain rule, the Liouville equation takes the multiscale form

$$\frac{\partial \rho}{\partial t} = \left(\mathcal{L}_0 + \varepsilon \mathcal{L}_1\right)\rho \tag{2.5}$$

$$\mathcal{L}_0 = -\sum_{i=1}^{N}\left(\frac{\vec{p}_i}{m_i}\bullet\frac{\partial}{\partial \vec{r}_i} + \vec{F}_i\bullet\frac{\partial}{\partial \vec{p}_i}\right) \tag{2.6}$$

$$\mathcal{L}_1 = -\sum_{l=1}^{M}\frac{\vec{P}_l}{m}\bullet\frac{\partial}{\partial \vec{R}_l}\,. \tag{2.7}$$

Introducing $\underline{R}$ -dependence in $\rho$ does not imply a change in the number of degrees of freedom (i.e., *6N*). Instead, the inclusion of these CM variables in the formulation is to acknowledge the multiple ways in which $\rho$ depends on the all-atom state $\Gamma$ . It has been assumed that each nanocomponent has total mass $m\varepsilon^{-1}$ , where *m* is a typical atomic mass. The operators $\mathcal{L}_0$ and $\mathcal{L}_1$ follow from the chain rule and the assumed dependence $\rho(\Gamma,\underline{R},t)$ of the *N*-atom probability density at time *t*.

A discrete multiscale development proceeds by introducing a short time variable $t_0 = t$ that tracks the $10^{-14}$ s atomistic fluctuations and a set of long-time variables $t_n = \varepsilon^n t; n = 1,2,\cdots$, used to track the distinct ways in which $\rho$ depends on time. With this, $\rho$ is cast as a power series in $\varepsilon$ and determined via a perturbative solution of the Liouville equation.

The lowest-order solution $\rho_0$ is assumed to be slowly varying (i.e., independent of $t_0$ ) and hence $\mathcal{L}_0\rho_0 = 0$ . Thus, $\rho_0$ is in the nullspace of $\mathcal{L}_0$ , and in particular, is determined using entropy maximization[12-16]. For the present problem, we obtain

$$\rho_0 = \frac{e^{-\beta H}}{Q} W\left(\underline{R},\underline{t}\right) \tag{2.8}$$

$$Q = \int d^{6N}\Gamma^* \Delta\left(\underline{R} - \underline{R}^*\right) e^{-\beta H^*} \tag{2.9}$$

where $H$ is the Hamiltonian generating $\mathcal{L}_0$. This lowest-order solution is developed using an information theory approach to construct the isothermal quasi-equilibrium ensemble for fixed values of $\underline{R}$. The development proceeds via an order-by-order analysis, eventually leading to a closed equation for $\Psi$, after inserting $\rho$ to $O(\varepsilon)$ in (2.3) and recognizing that $\Psi \to W$ as $\varepsilon \to 0$.

Unfortunately, this approach does not reveal the broad, essentially continuous, spectrum of timescales in the many-component system. For example, clusters of nanocomponents of increasingly large size are expected to evolve on increasingly long timescales. As *M* and *N* are very large for problems of interest, the development should reconstruct the $t_n$-dependence of $\rho$ for an extremely large set of *n*, while the classical, discrete timescale analysis ends at $t_2$. Thus we reformulate the problem in terms of a set of mode coordinates in favor of the component CM ($\underline{R}$) description.

Consider the method of Jaqaman and Ortoleva[17], as modified to construct order parameters for the multiscale analysis of *N*-atoms systems[13,15,16,19]. In this formulation, the position $\vec{r}_i$ of atom *i* is written as a sum of coherent and residual parts. The residual displacement $\vec{\sigma}_i$ accounts for fluctuations over-and-above coherent deformation. Hence the position of atom *i* is written in terms of a deformation from a reference position $\vec{r}_i^{\,0}$ using order parameters $\vec{\Phi}_k$, and basis functions $U_k$ via

$$\vec{r}_i = \sum_k \vec{\Phi}_k U_k\left(\vec{r}_i^{\,0}\right) + \vec{\sigma}_i . \tag{2.10}$$

Define a mass-weighted inner product $(a,b)$ via

$$(a,b) = \sum_{i=1}^{N} a_i b_i m_i . \tag{2.11}$$

If $a_i b_i = 1$ for all *i*, then $(a,b) = m_{tot}$, the total mass of the system. The $U_k$ are constructed to be orthogonal and normalized such that

$$\sum_{i=1}^{N} m_i U_k\left(\vec{r}_i^{\,0}\right) U_l\left(\vec{r}_i^{\,0}\right) = (U_k, U_l) = \mu_k \delta_{kl} , \tag{2.12}$$

where $\mu_k$ is a constant chosen to be an effective mass associated with the *k*-th order parameter. The label of the basis functions is chosen such that $\mu_k$ decreases with $k$. For example, if $k$ represents an inverse length characteristic of the spatial variation of $U_k$, then in one period of $U_k$, there is a volume of about $k^{-3}$ and hence a mass of similar magnitude. Thus, we suggest that $\mu_k$ behaves like $k^{-3}$ for large *k* since $\mu_k$ is $m_{tot}$ for small $k$ and about the mass of a single nanocomponent $m_c$ for the maximum cutoff value $k_c$. Therefore, $m_c \lesssim \mu_k \lesssim m_{tot}$. With *k* having the character of a wavevector for the *k*-th mode, we suggest that $\mu_k$ is qualitatively of the form $\mu_k = A + Bk^{-3}$ with *A* and *B* determined by imposing $\mu_k = m_{tot}$ when $k = 1$ and $\mu_k = m_c$ when $k = k_c$.

Upon minimizing the sum of the squares of residuals $|\vec{\sigma}_i|^2$ as in Pankavich et al[14-16], one finds

$$\vec{\Phi}_k = \frac{1}{\mu_k} \sum_{i=1}^{N} m_i U_k \left(\vec{r}_i^{\,0}\right) \vec{r}_i. \tag{2.13}$$

If $U_k$ is roughly one, then $\vec{\Phi}_k$ acts like the CM of the entire collection of components. For a more localized basis function, i.e., $\mu_k$ is roughly $m_c$, then $\vec{\Phi}_k$ acts like the CM of a component. We only choose the $U_k$ that vary on a length scale much greater than the average nearest-neighbor atomic spacing and, in particular, greater than the size of a nanocomponent, so that their effective mass is high.

When the structural components are fairly rigid, their internal dynamics are fast. Assuming the host fluid dynamics are also fast relative to the translation of the *M* units, a gap exists that separates the timescale of the nanocomponent dynamics from that of the fluctuations at the atomic scale. As above, we introduce a parameter $\varepsilon = m/m_c$ for typical atomic mass *m*. Using arguments as in earlier studies[11-16,18-19], the typical magnitude of the thermalized momentum of a nanocomponent is $\mathrm{O}\left(\varepsilon^{-1}\right)$. For constant $U_k$, $\vec{\Phi}_k$ is the CM of the entire collection of components. Thus, the $\vec{\Phi}_k$ evolve slowly when the associated basis function $U_k$ varies on a length scale greater than or equal to the size of a nanocomponent, i.e., when $m_c \lesssim \mu_k \lesssim m_{tot}$ for $1 \le k \le k_c$. If the number of components is $\mathrm{O}\left(\varepsilon^{-1}\right)$, then $m\varepsilon^{-1} \lesssim \mu_k \lesssim m\varepsilon^{-2}$ so that $\mu_k$ is large and varies greatly over the range of *k*. With this large range of effective mass, one expects there is an associated broad spectrum of timescales.

We postulate that for the phenomena of interest, two length scales exist on which units move: (1) the interatomic scale (i.e., as in close encounters between nanocomponents) that is characterized by $\Gamma$ , and (2) the long scale, $\mathrm{O}\left(\varepsilon^{-1}\right)$, movement that is related to larger scale reorganization (i.e., $\vec{\Phi}_k$ ). With $\vec{\Pi}_k \equiv -\mathcal{L}\vec{\Phi}_k$ , the chain rule implies that the Liouville equation takes the form $\mathcal{L}_0 + \hat{\mathcal{L}}$ where $\mathcal{L}_0$ is as in (2.6) and

$$\hat{\mathcal{L}} = -\sum_k \frac{1}{\mu_k} \frac{\vec{\Pi}_k}{m} \bullet \frac{\partial}{\partial \vec{\Phi}_k} . \tag{2.14}$$

The order parameters and momenta $\left\{\vec{\Phi}_k, \vec{\Pi}_k,\ k = 1, 2, \cdots, k_c\right\}$ are labeled by the integer *k*. However, if these order parameter disturbances are quasi-continuous, i.e., change smoothly in character with *k*, then we switch to labeling by a wavevector $\vec{k} \in \mathbb{R}^3$ with $k_m \le \left|\vec{k}\right| \le k_c$ and $k_m$ determined by the volume of the system. In developing the continuous scaling framework of the next section, we use the three-dimensional wavevector $\vec{k}$ to index order parameters and derive implications for the continuous scaling behavior of these large systems. Since we consider only systems with large volume, and hence include arbitrarily small wavevector lengths, we take $k_m = 0$ . Letting $\vec{k}$ vary over a continuum of values implies the existence of order parameters $\vec{\Phi}(\vec{k})$ , and conjugate momenta $\vec{\Pi}(\vec{k})$ , that depend continuously on the wavevector, in addition to the *N*-atom state $\Gamma$ .

## 3 A FUNCTIONAL DIFFERENTIAL SMOLUCHOWSKI EQUATION

The order parameter dynamics of the previous section naturally allow for the introduction of a continuous scaling and the derivation of associated Smoluchowski equations. In the proposed generalization of the discrete scaling of Sect. 2, we introduce the function $\omega(\vec{k})$ which increases monotonically with $k = |\vec{k}|$. In particular, we take $\omega(\vec{k}) \propto \mu_k^{-1}$ with $\mu_k$ as in Sect. 2. Since small $k$ (i.e., long wavelength) disturbances have the greatest effective mass $\mu_k$, these are the slowest excitations in the system. Recall from Set. 2 that the values of $k$ are restricted by the typical nanocomponent mass so that $k \le k_c$ throughout our analysis. In the following, we introduce a continuous set of $\vec{k}$-dependent order parameters, use $\omega(\vec{k})$ to order a perturbation expansion, and arrive at a scaling theory for systems with a broad continuum of timescales.

Order parameters for continuous scaling are introduced by writing atomic positions as an integral transform analogue of (2.8) via a continuum of basis functions $U(\vec{k}, \vec{r}_i^{\,0})$ which depend on the reference configuration $\vec{r}_i^{\,0}$:

$$\vec{r}_i = \frac{\Omega}{(2\pi)^3} \int_{k<k_c} d^3k \vec{\Phi}(\vec{k}) U(\vec{k}, \vec{r}_i^{\,0}) + \vec{\sigma}_i \,. \tag{3.1}$$

In transitioning from the discrete case to the continuum, we replace the sum on the integer index by a $\Omega/(2\pi)^3$-weighted integral, i.e., $\Omega d^3k/(2\pi)^3$ is the number of $\vec{k}$-index values in $d^3k$ for system volume $\Omega$. As for the discrete case, the $\vec{\sigma}_i$ represent the residual, "random" displacements from the coherent configuration generated by changes in the order parameters $\vec{\Phi}(\vec{k})$. For example, if $U(\vec{k}, \vec{r}_i^{\,0}) \propto \exp(i\vec{k} \bullet \vec{r}_i^{\,0})$, the order parameter $\vec{\Phi}(\vec{k})$ corresponds to the Fourier transform from $\vec{r}^{\,0}$-space to $\vec{r}$-space. Arguing by analogy with Sect. 2 and using (3.1), yields

$$\sum_{i=1}^{N} m_i U(\vec{k}, \vec{r}_i^{\,0}) \vec{r}_i = \frac{\Omega}{(2\pi)^3} \int_{|\vec{l}|<k_c} d^3l B(\vec{k}, \vec{l}) \vec{\Phi}(\vec{l}) \tag{3.2}$$

$$B(\vec{k}, \vec{l}) = \sum_{i=1}^{N} m_i U(\vec{k}, \vec{r}_i^{\,0}) U(\vec{l}, \vec{r}_i^{\,0}). \tag{3.3}$$

This constitutes a set of equations that determine the $\vec{\Phi}(\vec{k})$ in terms of the *N*-atom configuration. These expressions are obtained via minimization of the total mass-weighted square residual $R = \sum_{i=1}^{N} m_i \left( \vec{r}_i - \frac{\Omega}{(2\pi)^3} \int_{k<k_c} d^3k \vec{\Phi}\left(\vec{k}\right) U\left(\vec{k},\vec{r}_i^{\,0}\right) \right)^2$ over all order parameters $\vec{\Phi}\left(\vec{k}\right)$, i.e., by solving $\delta R / \delta \vec{\Phi}\left(\vec{k}\right) = \vec{0}$.

One may derive an equation for computing the OPs from Newtonian dynamics by taking $\mathcal{L}$ of both sides of (3.2). Letting $-\mathcal{L}\vec{\Phi}\left(\vec{k}\right) = \vec{\Pi}(\vec{k})$ yields

$$\int_{|\vec{l}|<k_c} d^3l B(\vec{k},\vec{l}) \vec{\Pi}(\vec{l}) = \sum_{i=1}^{N} U\left(\vec{k},\vec{r}_i^{\,0}\right) \vec{p}_i \tag{3.4}$$

which determine the $\vec{\Pi}(\vec{k})$ in terms of the *N*-atom configuration. Integral equations for $\vec{\Phi}\left(\vec{k}\right)$ and $\vec{\Pi}\left(\vec{k}\right)$ emerge which simplify greatly for orthogonal basis functions. For instance, if $\sum_{i=1}^{N} m_i U\left(\vec{k},\vec{r}_i^{\,0}\right) U\left(\vec{l},\vec{r}_i^{\,0}\right) = \mu(\vec{k}) \delta(\vec{k}-\vec{l})$ with $\mu(\vec{k}) = \mu_k$ from Sect. 2, then (3.2) becomes an explicit equation for the OPs:

$$\vec{\Phi}(\vec{k}) = \frac{1}{\mu(\vec{k})} \sum_{i=1}^{N} m_i U\left(\vec{k},\vec{r}_i^{\,0}\right) \vec{r}_i \ . \tag{3.5}$$

In this case, the expression for $\vec{\Pi}(\vec{k})$ simplifies as well. Introducing $\omega(\vec{k}) = 1/\mu(\vec{k})$, one obtains $-\mathcal{L}\vec{\Phi}\left(\vec{k}\right) = \omega(\vec{k}) \vec{\Pi}(\vec{k})$ where

$$\vec{\Pi}(\vec{k}) = \sum_{i=1}^{N} U\left(\vec{k},\vec{r}_i^{\,0}\right) \vec{p}_i \ . \tag{3.6}$$

We take (3.5) and (3.6) as the definitions of the OPs and their conjugate momenta, though variations will occur if basis functions do not satisfy mass-weighted orthogonality.

The reduced probability density $\Psi\left[\vec{\Phi},t\right]$ is a functional of $\vec{\Phi}$ and a function of $t$. In the discrete case, $\Psi$ took the form

$$\Psi(\underline{\Phi},t) = \int d^{6N}\Gamma^* \Delta(\vec{\Phi} - \vec{\Phi}^*)\rho \tag{3.7}$$

where $\Delta$ was a product of delta functions, one for each order parameter. Generalizing to the continuous case, we write the limit of this product of delta functions as

$$\Delta\left[\vec{\Phi} - \vec{\Phi}^*\right] = \exp\left[\frac{\Omega}{(2\pi)^3}\int d^3k \ln\delta\left(\vec{\Phi}\left(\vec{k}\right) - \vec{\Phi}^*\left(\vec{k}\right)\right)\right]. \tag{3.8}$$

Hence, $\Psi$ is represented in the same form as in Sect. 2, but with $\Delta$ having functional dependence on $\vec{\Phi}$ as defined by (3.8). Taking the time derivative of both sides of (3.7) and using the Liouville equation with the chain rule, one obtains the functional conservation law

$$\frac{\partial\Psi}{\partial t} = -\frac{\Omega}{(2\pi)^3}\int_{k<k_c} d^3k\,\omega\left(\vec{k}\right)\frac{\delta}{\delta\vec{\Phi}\left(\vec{k}\right)}\bullet\int d^{6N}\Gamma^*\Delta\left[\vec{\Phi} - \vec{\Phi}^*\right]\vec{\Pi}^*\left(\vec{k}\right)\rho\,. \tag{3.9}$$

We construct $\rho$ as a functional expansion in $\omega(\vec{k})$ using a generalization of the power series in $\varepsilon$ for the discrete case, and then use (3.9) to derive a closed equation for $\Psi$ (i.e, one that does not depend on the full *N*-atom probability density $\rho$, but only involves $\Psi$).

In traditional multiscale theory, one assumes that an *N*-atom system can be characterized as evolving on a discrete set of timescales, each separated from nearby timescales with a finite gap. The width of these gaps scales as an inverse power of a small parameter (e.g., the ratio of the mass of a typical atom to that of a many-atom nanocomponent). But the discussion at the end of Sect. 2 suggests that for many systems there is a myriad of timescales, each of which is separated from those nearby by such a small gap that the scaling is essentially continuous, i.e., the underlying physics changes in an essentially continuous manner across these scales. Therefore, we now investigate a theory wherein a finite gap, as in Fig. 1, separates a discrete timescale (i.e., that of rapidly-fluctuating atomistic processes) from a quasi-continuum of scales associated with a spectrum of slow processes. The presence of a continuum of functional Liouville-type operators on the right side of (3.9), suggests there is a related continuum of operators that correspond to an uncountable set of time variables on the left side as follows.

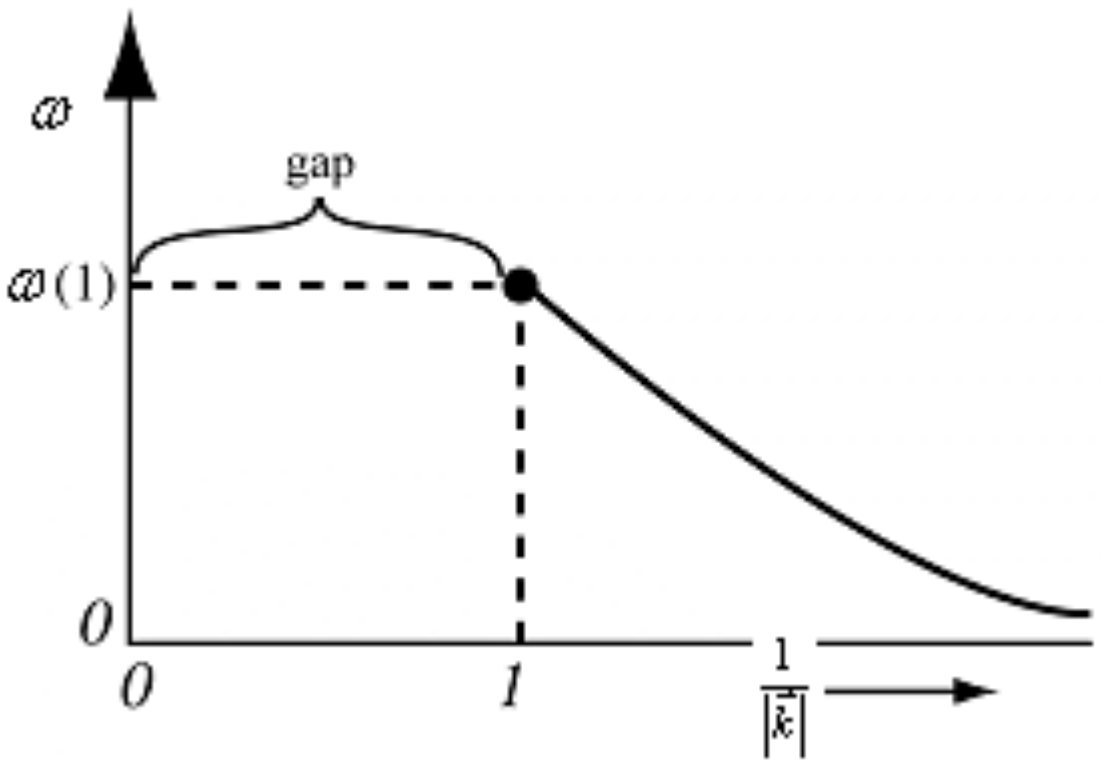

**Fig. 1** The graph of a continuous scaling factor for slow modes that operate on a distinctly longer timescale than that of rapid atomic-scale fluctuations is shown. The gap in the time scale of modes that are active is created by the nature of the initial data.

In analogy to the discrete case, we make the ansatz $\rho = \rho\left[\Gamma, \vec{\Phi}, t\right]$, so that $\rho$ has an explicit functional dependence on the continuum of OPs, in addition to its explicit $\Gamma$ -dependence. The Liouville operator $\mathcal{L}$ is then written as a sum of discrete and continuously scaled contributions:

$$\frac{\partial \rho}{\partial t} = \mathcal{L}\rho \tag{3.10}$$

$$\mathcal{L} = \mathcal{L}_0 + \hat{\mathcal{L}} \tag{3.11}$$

With these definitions and the chain rule, we obtain an explicit form for the $\hat{\mathcal{L}}$ operator:

$$\hat{\mathcal{L}} = -\frac{\Omega}{(2\pi)^3} \int_{k<k_c} d^3k\omega\left(\vec{k}\right)\vec{\Pi}(\vec{k}) \bullet \frac{\delta}{\delta\vec{\Phi}(\vec{k})}. \tag{3.12}$$

In the following, we treat the set of short scales as discrete and isolated. The fast timescale dynamics are driven by $\mathcal{L}_0$, accounting for atomic-scale processes whose timescale is that of atomic collision or vibration (i.e., $10^{-14}$ seconds). In contrast, $\hat{\mathcal{L}}$ accounts for the long-length scale, slow dynamics of the order parameters, and thus involves functional derivatives with respect to the continuous family of order parameters $\vec{\Phi}\left(\vec{k}\right)$.

In analogy with traditional discrete multiscale treatments of the Liouville equation[11-16], we introduce a continuous spectrum of time variables to capture the effects of $\hat{\mathcal{L}}$. As a consequence, the reformulation develops a continuum character, i.e., traditional sums over a discrete timescale index are replaced by a $\vec{k}$ integration and derivatives are replaced by functional derivatives. To achieve self-consistency, we first introduce times $\tau_0 \equiv t$, $\tau(\vec{k}) = \omega(\vec{k})t; k \le k_c$. The short time variable $\tau_0$ is the analogue of $t_0$ in Sect. 2, while the continuum of times $\tau(\vec{k})$ is the analogue of the traditional discrete sequence $t_1 = \varepsilon t,\ t_2 = \varepsilon^2 t, \cdots$. It is thereby assumed that $\rho$ depends on the discrete short time $\tau_0 = t$ and the uncountable continuum of long times $\left\{\tau\left(\vec{k}\right); k \le k_c\right\}$.

In this formulation, the $N$-atom probability density is a function of $\Gamma$, the set of $N$ atomic positions and momenta, as well as, the short time variable $\tau_0$, and a functional of a continuous set of order parameters $\vec{\Phi}$ and long time variables $\tau(\vec{k})$. Through the chain rule, $\mathcal{L}_0$ operates on the explicit $\Gamma$-dependence, while the continuum term $\hat{\mathcal{L}}$ operates on the indirect functional dependence, through $\vec{\Phi}$, and drives the slow dynamics as implied by the smallness of the scaling function $\omega\left(\vec{k}\right)$ over its range of values. The scaling function $\omega\left(\vec{k}\right)$ decreases to zero as $k = \left|\vec{k}\right| \to 0$, reflecting the fact that longer wavelength (smaller $k$) modes are slower varying. Since $\rho$ has direct dependence on the microscopic variables $\Gamma = \left\{\vec{r}_i, \vec{p}_i; i = 1, \cdots, N\right\}$ and indirect dependence on them through the continuum of variables $\left\{\vec{\Phi}(\vec{k}); k \le k_c\right\}$, the multiscale Liouville equation becomes a mixed differential - functional differential equation. The weight $\omega(\vec{k})$ plays the role of the scaling factor $\varepsilon$ raised to various discrete powers. For example, $\omega(\vec{k}) = \varepsilon^{k_c/k}$ for $\varepsilon \ll 1$ and $0 < k \le k_c$ is an example of a scaling function.

The above formulation is not a violation of the number of degrees of freedom, as the $6N$ degrees of freedom constituting $\Gamma$ already provide a complete description of the state of a classical system. However, it was shown for discrete scaling systems[12-16] that the dual dependence of $\rho$ on $\Gamma$ and OPs is simply a reflection of the multiple ways that $\rho$ depends on the $N$-atom configuration. Thus, $\rho$ depends on $\Gamma$ both directly and through the order parameters, indirectly.

The chain rule and the functional dependence of $\rho$ on the OPs and the continuum of time variables imply that (3.10) becomes

$$\left(\frac{\partial}{\partial \tau_0} + \frac{\Omega}{(2\pi)^3} \int_{k<k_c} d^3k\,\omega\left(\vec{k}\right) \frac{\delta}{\delta \tau\left(\vec{k}\right)} + \cdots \right)\rho = \left(\mathcal{L}_0 + \frac{\Omega}{(2\pi)^3} \int_{k<k_c} d^3k\,\omega\left(\vec{k}\right) \vec{\Pi}(\vec{k}) \cdot \frac{\delta}{\delta \vec{\Phi}(\vec{k})}\right)\rho . \qquad (3.13)$$

With these arguments, the challenge is to demonstrate that (1) this is a self-consistent mathematical framework, (2) a perturbation analysis can be performed, and (3) a generalized functional Smoluchowski equation can be developed that captures the long-time, stochastic dynamics of systems like a large collection of nanoparticles.

Since $\omega(\vec{k})$ is small (reflecting the long timescales involved), a Volterra-type functional expansion in $\omega$ is introduced for $\rho$:

$$\rho = \rho_0 + \sum_{\ell=1}^{\infty} \tfrac{1}{\ell!} \frac{\Omega^{\ell}}{(2\pi)^{3\ell}} \int_{k<k_c} d^3k_1 \cdots d^3k_\ell \, \omega(\vec{k}_1) \cdots \omega(\vec{k}_\ell) \rho_{\vec{k}_1} \cdots \rho_{\vec{k}_\ell} . \tag{3.14}$$

The missing terms indicated on the LHS of (3.13) arise from the need to introduce additional types of time variables (e.g. $\tau(\vec{k}_1,\vec{k}_2) = \omega(\vec{k}_1)\omega(\vec{k}_2)t$). These higher-order time variables will arise because, in general, there is no simple relation between the times corresponding to differing wavevector values (e.g. $\omega(\vec{k}_1)\tau(\vec{k}_2) = \tau(\vec{k}_1,\vec{k}_2)$). However, since our analysis need only be carried out to first order in the functional expansion of $\rho$, we will not elucidate further. Eqn (3.13) is then solved using (3.14) in an "order-by-order" fashion, i.e., in a manner analogous to Sect. 2 in which the equations were solved in powers of $\varepsilon$. Using information theory to construct the lowest-order solution and the Gibbs-hypothesized equivalence of thermal and long-time averages to remove secular behavior in the solution at each order, we obtain expressions for $\rho_0$ and the functions $\rho_{\vec{k}_1}$ thru $\rho_{\vec{k}_\ell}$. Upon recomposing $\rho$ and using these expressions in (3.9), we find

$$\frac{\partial \Psi}{\partial t} = \frac{\Omega^2}{(2\pi)^6} \int_{k_1<k_c} \int_{k_2<k_c} d^3k_1 d^3k_2 \omega(\vec{k}_1)\omega(\vec{k}_2) \frac{\delta}{\delta\Phi(\vec{k}_1)} \cdot \left[ \bar{\bar{D}} \left( -\beta \vec{f}^{th}(\vec{k}_2) + \frac{\delta}{\delta\Phi(\vec{k}_2)} \right) \Psi \right] \tag{3.15}$$

$$\bar{\bar{D}}(\vec{k}_1,\vec{k}_2) = \int_{-\infty}^{0} dt'' \left[ \vec{\Pi}(\vec{k}_1) e^{-\mathcal{L}_0 t''} \vec{\Pi}(\vec{k}_2) \right]^{th} \tag{3.16}$$

We term this a functional Smoluchowski equation for continuous scaling systems. Here, the $\vec{f}^{th}$ terms represent thermal-averaged forces, which can be computed using an order-parameter-constrained ensemble of atomistic configurations. Details of these calculations are provided in the appendix.

## 4. THE NONLOCAL LANGEVIN EQUATION

Since the Smoluchowski equation of the previous section cannot be solved analytically or numerically due to practical difficulties, the equivalent Langevin equations are the only practical way to simulate the stochastic processes of interest here. We now show that this Langevin equation is nonlocal and describes the stochastic dynamics of the continuum of order parameters $\vec{\Phi}$. We prove the Monte Carlo-equivalence by postulating the form of the Langevin equation and comparing the resulting Smoluchowski equation with that derived in Sect. 3.

Our postulated form of the Langevin equation is

$$\frac{\partial \vec{\Phi}(t,\vec{k})}{\partial t} = \int d^3 l \vec{\vec{\alpha}}(\vec{k},\vec{l}) \left\langle \vec{f}(\vec{l}) \right\rangle + \vec{\xi}(t,\vec{k}) \tag{4.1}$$

for $k \le k_c$. Here, $\vec{\vec{\alpha}}(\vec{k},\vec{l})$ is a diffusion kernel that we will relate to the diffusion coefficients $\vec{\vec{D}}$ of (3.16), $\left\langle \vec{f}(\vec{l}) \right\rangle$ is the thermal-averaged force for the $\vec{l}$ -th order parameter, and $\vec{\xi}$ is a random force driving the stochastic dynamics of $\vec{\Phi}(t,\vec{k})$. As described at the end of the previous section, the thermal-averaged forces are defined via an OP-constrained ensemble, and hence depend only upon the order parameters in the system. We now determine conditions under which (4.1) is Monte Carlo-equivalent to (3.15) with (3.16). For simplicity, we have introduced units for the wavevector such that the $\Omega/(2\pi)^3$ factors do not appear.

Let $\Psi\left[\vec{\Phi},t\right]$ be the probability density functional for the continuum of order parameters $\vec{\Phi}$. We assume the set of random forces $\vec{\xi}$ change much more rapidly over time than $\vec{\Phi}$. A time interval $\Delta t$ is postulated to exist such that the system experiences a representative sample of variations in $\vec{\xi}\left(t,\vec{k}\right)$ during $\Delta t$, and yet $\vec{\Phi}(t,\vec{k})$ changes only a small amount during $\Delta t$ due to $\left\langle \vec{f}(\vec{l}) \right\rangle$. We adopt a Markov assumption such that $\Psi$ can be advanced from $t$ to $t+\Delta t$ solely knowing $\Psi$ at time $t$ and the transition probability for a change in $\vec{\Phi}(t,\vec{k})$ during $\Delta t$. The evolution of $\Psi$ is thus determined by the statistics of all $\vec{\xi}\left(t,\vec{k}\right)$ time courses. Let $T\left[\vec{\xi},t,\Delta t\right]$ be the probability functional for a scenario of $\vec{\xi}$ during $t$ to $t+\Delta t$. Then, $\Psi$ evolves via the Markov expression

$$\Psi\left[\vec{\Phi},t+\Delta t\right] = \underset{\vec{\Phi}'}{\mathcal{S}} \underset{\vec{\xi}}{\mathcal{S}} T\left[\vec{\xi},t,\Delta t\right] \Delta\left[\vec{\Phi} - \vec{\tilde{\Phi}}\left[\vec{\xi};t,\Delta t,\vec{\Phi}'\right]\right] \Psi\left[\vec{\Phi}',t\right]. \tag{4.2}$$

Here, $\bar{\bar{\Phi}}$ is the solution of (4.1) at time $t+\Delta t$, given that the state was $\bar{\Phi}'$ at $t$ and $\Delta$ is defined by (3.8). The functional integrals in (4.2) are over all order parameter states $\bar{\Phi}'$ (for all $\vec{k}$) and scenarios of $\bar{\xi}$ for times between $t$ and $t+\Delta t$.

A Volterra series expansion of $\Delta$ in the changes $\bar{\Phi}'-\bar{\bar{\Phi}}$ (that vanish as $\Delta t \to 0$) takes the form

$$\Delta\left(\bar{\Phi}-\bar{\bar{\Phi}}\right) \approx \Delta\left(\bar{\Phi}-\bar{\Phi}'\right)+\sum_{\ell=1}^{\infty}\tfrac{1}{\ell!}\int d^3k_1\cdots d^3k_\ell\left(\bar{\Phi}'-\bar{\bar{\Phi}}\right)(\vec{k}_1)\cdots\left(\bar{\Phi}'-\bar{\bar{\Phi}}\right)(\vec{k}_\ell)\frac{\delta^\ell\Delta\left(\bar{\Phi}-\bar{\Phi}'\right)}{\delta\bar{\Phi}(\vec{k}_1)\cdots\delta\bar{\Phi}(\vec{k}_\ell)}. \tag{4.3}$$

Denote the averaging implied by the $T$-weighted functional integrals in (4.2) as $\langle\ \rangle_T$. Under the postulate that $\Psi$ changes only a small amount over $\Delta t$, one finds

$$\Psi\left[\bar{\Phi},t+\Delta t\right]\approx\Psi\left[\bar{\Phi},t\right]+\frac{\partial\Psi}{\partial t}\Delta t. \tag{4.4}$$

Substituting this and (4.3) to second order into (4.2), and neglecting terms of order higher than $\Delta t$ yields

$$\begin{aligned}\frac{\partial\Psi}{\partial t}\Delta t &= \int d^3k_1\frac{\delta}{\delta\bar{\Phi}\left(\vec{k}_1\right)}\bullet\left\{\left\langle\left(\bar{\Phi}-\bar{\bar{\Phi}}\right)\left(\vec{k}_1\right)\right\rangle_T\Psi\right] \\ &+\frac{1}{2}\int\int d^3k_1d^3k_2\frac{\delta^2}{\delta\bar{\Phi}\left(\vec{k}_1\right)\delta\bar{\Phi}\left(\vec{k}_2\right)}\bullet\left\{\left\langle\left(\bar{\Phi}-\bar{\bar{\Phi}}\right)\left(\vec{k}_1\right)\left(\bar{\Phi}-\bar{\bar{\Phi}}\right)\left(\vec{k}_2\right)\right\rangle_T\Psi\right\}\end{aligned} \tag{4.5}$$

To better represent the $\bar{\bar{\alpha}}$-kernel, define the linear operator $P$ by

$$\left(\bar{P}\bar{\Psi}\right)(\vec{k})=\int d^3l\,\bar{\bar{\alpha}}(\vec{k},\vec{l})\bar{\Psi}(\vec{l}) \tag{4.6}$$

for any integrable function $\bar{\Psi}$. Integration of (4.1) yields

$$\bar{\bar{\Phi}}\left(\vec{k}\right)=\bar{\Phi}\left(\vec{k}\right)+\tau\left(\bar{P}\langle\bar{f}\rangle\right)\left(\vec{k}\right)+\int_t^{t+\Delta t}dt'\bar{\xi}(t',\vec{k}). \tag{4.7}$$

To derive formulas for the $\bar{\Phi}-\bar{\bar{\Phi}}$ terms in (4.5), we assume the average random force vanishes for every $\vec{k}$, that is, $\left\langle\bar{\xi}(t,\vec{k})\right\rangle_T=\vec{0}$. If the process generating $\bar{\xi}(t,\vec{k})$ is stationary, then $\left\langle\bar{\xi}(t,\vec{k}_1)\bar{\xi}(t',\vec{k}_2)\right\rangle_T$ only depends on $t-t'$, and not on $t$ and $t'$ independently, i.e, $\left\langle\bar{\xi}(t,\vec{k}_1)\bar{\xi}(t',\vec{k}_2)\right\rangle_T\equiv\bar{\bar{\phi}}\left(t-t',\vec{k}_1,\vec{k}_2\right)$.

Neglecting terms of order higher than $\Delta t$, the statistical properties of $\bar{\xi}$ and (4.7) imply

$$\left\langle \vec{\Phi}\left(\vec{k}\right) - \bar{\vec{\Phi}}\left(\vec{k}\right)\right\rangle_T = -\left(\vec{P}\langle\vec{f}\rangle\right)\left(\vec{k}\right)\Delta t \tag{4.8}$$

$$\left\langle \left(\vec{\Phi} - \bar{\vec{\Phi}}\right)\left(\vec{k}_1\right)\left(\vec{\Phi} - \bar{\vec{\Phi}}\right)\left(\vec{k}_2\right)\right\rangle_T = \int_0^{\Delta t} dt' \int_{-t'}^{\Delta t - t'} ds \bar{\vec{\phi}}\left(s, \vec{k}_1, \vec{k}_2\right). \tag{4.9}$$

For essentially all $t'$ in the time period $(0, \Delta t)$, the upper limit of the $s$ integral is much greater than the correlation time. Since $\phi(|-t'|)$ is negligible except for $t' << \Delta t$, one finds

$$\left\langle \left(\vec{\Phi} - \bar{\vec{\Phi}}\right)\left(\vec{k}_1\right)\left(\vec{\Phi} - \bar{\vec{\Phi}}\right)\left(\vec{k}_2\right)\right\rangle_T = \bar{\vec{A}}(\vec{k}_1, \vec{k}_2)\Delta t \tag{4.10}$$

where

$$\bar{\vec{A}}(\vec{k}_1, \vec{k}_2) \equiv \int_{-\infty}^{0} ds \bar{\vec{\phi}}\left(s, \vec{k}_1, \vec{k}_2\right). \tag{4.11}$$

These values of $\left\langle \left(\vec{\Phi}\left(\vec{k}\right) - \bar{\vec{\Phi}}\left(\vec{k}\right)\right)\right\rangle_T$ and $\left\langle \left(\vec{\Phi} - \bar{\vec{\Phi}}\right)\left(\vec{k}_1\right)\left(\vec{\Phi} - \bar{\vec{\Phi}}\right)\left(\vec{k}_2\right)\right\rangle_T$ are inserted into (4.5) to yield the functional Smoluchowski equation:

$$\frac{\partial \Psi}{\partial t} = \int\int d^3k_1 d^3k_2 \frac{\delta}{\delta \vec{\Phi}\left(\vec{k}_1\right)} \cdot \left[\bar{\vec{\alpha}}\left(\vec{k}_1, \vec{k}_2\right)\left(\frac{1}{2}\left(\bar{\vec{\alpha}}^{-1}\left(\vec{k}_1, \vec{k}_2\right)\bar{\vec{A}}\left(\vec{k}_1, \vec{k}_2\right)\right)\frac{\delta}{\delta \vec{\Phi}\left(\vec{k}_2\right)} - \left\langle \vec{f}\left(\vec{k}_2\right)\right\rangle\right)\Psi\right]. \tag{4.12}$$

At long times, the closed, isothermal system reaches equilibrium. Hence $\Psi\left[\vec{\Phi}, t\right] \underset{t\to\infty}{\sim} \frac{\exp(-\beta F)}{Z} \equiv \Psi^{eq}\left[\vec{\Phi}\right]$ where $Z = \underset{\vec{\Phi}}{\mathcal{S}} e^{-\beta F}$. As this must be a time-independent solution of the functional Smoluchowski equation, we find $\frac{\beta}{2}\left(\bar{\vec{\alpha}}^{-1}\bar{\vec{A}}\right)(\vec{k}_1, \vec{k}_2) = \mathbb{I}$. Using (4.11) this relation constrains the statistics of $\vec{\xi}$ by

$$\frac{\beta}{2}\int_{-\infty}^{0} ds \left\langle \vec{\xi}(0, \vec{k}_1)\vec{\xi}(s, \vec{k}_2)\right\rangle_T = \bar{\vec{\alpha}}(\vec{k}_1, \vec{k}_2) \tag{4.13}$$

for every $\vec{k}_1, \vec{k}_2$. Comparison of the functional Smoluchowski equation (3.15) with (4.12), derived from the postulated Langevin equation, implies a relation between the diffusion coefficients $\bar{\vec{D}}$ and the kernel $\bar{\vec{\alpha}}$, namely

$$\bar{\vec{\alpha}}(\vec{k}_1, \vec{k}_2) = \beta \bar{\vec{D}}(\vec{k}_1, \vec{k}_2) \tag{4.14}$$

for every $\vec{k}_1, \vec{k}_2$. Combining (4.13) and (4.14), we can relate the statistics of $\vec{\xi}$ to $\bar{\bar{D}}$. Thus there is a well-defined relationship between the integrated, fluctuating force correlation functions and the diffusion coefficients $\bar{\bar{D}}$, a classic fluctuation-dissipation relation. The Langevin description can then be used to simulate many-component systems using $\vec{k}$ -space finite element techniques and Monte Carlo sampling of the random force. An algorithm for co-evolving the OPs and quasi-equilibrium, atomic probability density can be constructed by advancing the OPs at each Langevin timestep and simultaneously reconstructing the thermal-averaged forces and diffusion coefficients, allowing for the next advancement of the OPs in the implicit Langevin dynamics. This will avoid the need to impose simple phenomenological expressions for the thermal-averaged forces that may miss the interscale feedback between atomistic variables and OPs.

## 5. CONCLUSIONS

The broad quasi-continuum of characteristic times displayed by a system of many interacting nanocomponents is but one example of physical systems for which one cannot use the paradigm of a few timescales separated by large gaps. This latter classic approach has dominated multiscale analysis. We have presented a novel approach wherein a continuum of characteristic lengths and times, and associated variables, enables the derivation of a functional generalization of the Smoluchowski equation. This equation provides a framework for understanding the stochastic dynamics of order parameter field variables.

This multiscale approach, though similar to the projection operator method, contains subtle and important differences. The projection operator technique starts with the identification of descriptive variables of interest and, like the multiscale procedure, derives equations for the reduced probability density for these variables. However, the former technique introduces memory kernels; if these kernels have long-time memory then for the large systems of interest they cannot practically be evaluated using computer molecular dynamics. Furthermore, the kernels involve a propagator which is generated by a modified Liouville operator that cannot be constructed without pre-determining the reduced probability. If the kernels only have short-time memory, then the projection operator technique indeed reduces to the multiscale formalism. Since the latter is more readily generalized to a variety of problems, and the associated computations lead more directly to the reduced equation without involving a heuristic assumption on the characteristics of the memory kernel, the present methodology is preferable for systems with short memory. Key aspects of our formalism are the following:

- a set of order parameter field variables with an associated effective mass central to the scaling of their slow behavior
- an uncountable set of time variables related to laboratory time via a wavevector-dependent scaling function
- the ansatz that the *N*-atom probability density can be viewed as depending on the *N*-atom state $\Gamma$ both directly and, via a continuum of wavevector-dependent order parameters, indirectly

- a functional expansion in the inverse effective mass with the consequent emergence of a Smoluchowski equation, and Monte Carlo-equivalent Langevin equation, for the dynamics of order parameter field variables.

The theory has application to the self-assembly and dynamics of composite materials from nanocomponents, self-organization in cellular systems, flow in complex media, and, more generally, systems of many strongly interacting components evolving over a broad spectrum of timescales.

## APPENDIX: MULTISCALE ANALYSIS AND CONTINUOUS TIME SCALING

To further explore the analysis of Sect. 3, we begin with the ansatz on the probability density $\rho = \rho\left[\Gamma, t; \vec{\Phi}\right]$, so that $\rho$ depends on $\Gamma$ implicitly via the continuum of OPs in addition to its explicit $\Gamma$ -dependence. We rescale the wavevectors by letting $\vec{k}' = \Omega^{1/3}\vec{k}/2\pi$ and then drop the prime for brevity. We take the $\vec{k}$ integrals which appear below to be over the sphere $k \le k_c$ for cutoff $k_c$ where $k = \left|\vec{k}\right|$. The Liouville operator $\mathcal{L}$ can then be written (via use of the chain rule) as a sum of discrete and continuously scaled contributions:

$$\frac{\partial \rho}{\partial t} = \mathcal{L}\rho \tag{A.1}$$

$$\mathcal{L} = \mathcal{L}_0 + \hat{\mathcal{L}} \tag{A.2}$$

$$\hat{\mathcal{L}} = -\int d^3k\,\omega\left(\vec{k}\right)\vec{\Pi}(\vec{k})\bullet\frac{\delta}{\delta\vec{\Phi}(\vec{k})}. \tag{A.3}$$

In this formulation, the *N*-atom probability density is a function of $\Gamma$, the set of *N* atomic positions and momenta, and a functional of a continuous set of order parameters. Hence, $\mathcal{L}_0$ operates on the former while the $\hat{\mathcal{L}}$ term operates on the latter, and the Liouville equation becomes a mixed differential/functional-differential equation.

As discussed in Sect. 3, we introduce a continuous spectrum of time variables $\tau_0 \equiv t,\ \tau(\vec{k}) = \omega(\vec{k})t$ to capture the effects of $\mathcal{L}_0$ and $\hat{\mathcal{L}}$ respectively. Hence, $\rho$ is a function of $\tau_0$ and the *N*-atom state $\Gamma$ , while it is a functional of the slow times $\tau$ and $\vec{\Phi}$ . The chain rule and the functional dependence of $\rho$ imply that (3.1) becomes

$$\left(\frac{\partial}{\partial \tau_0} + \int\limits_{k\le k_c} d^3k\,\omega\left(\vec{k}\right)\frac{\delta}{\delta\tau\left(\vec{k}\right)} + \cdots\right)\rho = \left(\mathcal{L}_0 + \int\limits_{k\le k_c} d^3k\,\omega\left(\vec{k}\right)\vec{\Pi}(\vec{k})\bullet\frac{\delta}{\delta\vec{\Phi}(\vec{k})}\right)\rho\,. \tag{A.4}$$

While the set of times $\tau\left(\vec{k}\right)$ is satisfactory for the derivation of the Smoluchowski equation presented here, the fact that $\omega\left(\vec{k}_1\right), \omega\left(\vec{k}_2\right)$ cannot be related to $\omega\left(\vec{k}_3\right)$, for $\vec{k}_3$ related to $\vec{k}_1, \vec{k}_2$ implies the need to introduce more complex time variables (e.g., $\tau\left(\vec{k}_1, \vec{k}_2\right)$) as suggested by the omitted terms in (A.4).

Since $\omega(\vec{k})$ is small, a Volterra functional expansion in $\omega$ is introduced for $\rho$ :

$$\rho = \rho_0 + \sum_{\ell=1}^{\infty} \tfrac{1}{\ell!} \int_{k<k_c} d^3k_1 \cdots d^3k_\ell \, \omega(\vec{k}_1) \cdots \omega(\vec{k}_\ell) \rho_{\vec{k}_1} \cdots \rho_{\vec{k}_\ell} \, . \tag{A.5}$$

To lowest order in $\omega$ the Liouville equation implies

$$\frac{\partial \rho_0}{\partial \tau_0} = \mathcal{L}_0 \rho_0 \, . \tag{A.6}$$

For quasi-equilibrium states $\rho_0$ is independent of $\tau_0$, but it does depend on the slow times $\tau$. Hence, (A.6) becomes

$$\mathcal{L}_0 \rho_0 = 0 \, . \tag{A.7}$$

The form of $\rho_0$ is developed in analogy to the nanocanonical ensemble approach[12-16]:

$$\rho_0 = \hat{\rho} W, \quad \hat{\rho} = \frac{e^{-\beta H}}{\Xi}, \tag{A.8}$$

where $\Xi$ is the partition function

$$\Xi = \int d^{6N}\Gamma^* \Delta\left(\vec{\Phi} - \vec{\Phi}^*\right) e^{-\beta H^*} \tag{A.9}$$

and $W\left[\vec{\Phi}, \tau\right]$ is as yet unknown. Here, $\Delta$ is defined by (3.8), and $\vec{\Phi}^*$ represents the value of the order parameter field in terms of the integrated atomic variables. The functional derivative of $\Xi$ is related to the thermal-averaged force, i.e. $\delta(\ln \Xi) / \delta \vec{\Phi}(\vec{k}) = -\beta \vec{f}^{th}(\vec{k})$.

The analysis proceeds by combining (A.4) and (A.5), taking the $\ell$-th functional derivative of both sides of the result with respect to $\omega$, and setting $\omega$ to zero for $\ell = 1, 2, \cdots$. To first order one obtains

$$\frac{\delta \rho_0}{\delta \tau(\vec{k})} + \frac{\partial \rho_{\vec{k}_1}}{\partial \tau_0} = \mathcal{L}_0 \rho_{\vec{k}_1} + \vec{\Pi}(\vec{k}) \cdot \frac{\delta \rho_0}{\delta \vec{\Phi}(\vec{k})} \, . \tag{A.10}$$

Assuming the system begins in a quasi-equilibrium state given by $\rho_0$, i.e. $\rho_{\vec{k}_1}$ vanishes at $\tau_0 = 0$, we obtain

$$\rho_{\vec{k}_1} = \int_0^{\tau^*} dt' e^{\mathcal{L}_0(\tau_0 - t')} \left[ -\frac{\delta \rho_0}{\delta \tau(\vec{k})} + \vec{\Pi}(\vec{k}) \bullet \frac{\delta \rho_0}{\delta \vec{\Phi}(\vec{k})} \right]. \tag{A.11}$$

Changing variables such that $t'' = \tau_0 - t'$ yields

$$\rho_{\vec{k}_1} = -\tau_0 \hat{\rho} \frac{\delta W}{\delta \tau(\vec{k})} + \int_{-\tau_0}^{0} dt'' e^{-\mathcal{L}_0 t''} \vec{\Pi}(\vec{k}) \bullet \frac{\delta \rho_0}{\delta \vec{\Phi}(\vec{k})} . \tag{A.12}$$

Using the definition of $\hat{\rho}$, we find the integral of the last term in (A.12) takes the form $\vec{\Pi}(\vec{k}) \hat{\rho} \left[ -\beta \vec{f}^{th}(\vec{k}) W + \frac{\delta W}{\delta \vec{\Phi}(\vec{k})} \right]$. To remove secular behavior, we utilize the fundamental postulate of equilibrium statistical mechanics:

$$A^{th} = \lim_{\tau_0 \to \infty} \frac{1}{\tau_0} \int_{-\tau_0}^{0} dt'' e^{-\mathcal{L}_0 t''} A, \tag{A.13}$$

i.e., the thermal and long-time averages of any quantity are equal. The order parameters evolve slowly as compared to the $10^{-14}$ second atomistic timescale. Hence, we expect that the momenta come to a slowly evolving equilibrium, i.e., their ensemble average as weighted by $\hat{\rho}$ is zero. Using (A.13), it follows that the long time average of $\vec{\Pi}$ is zero. Upon dividing by $\tau_0$, taking the limit as $\tau_0$ tends to infinity, and using (A.13) in (A.12), we obtain

$$\frac{\delta W}{\delta \tau(\vec{k})} = 0 . \tag{A.14}$$

Thus, $\delta W / \delta \tau(\vec{k})$ vanishes and $W$ is independent of the long times $\tau(\vec{k})$. Hence, (A.12) becomes

$$\rho_{\vec{k}_1} = \hat{\rho} \int_{-\tau_0}^{0} dt'' e^{-\mathcal{L}_0 t''} \vec{\Pi}(\vec{k}) \left[ -\beta \vec{f}^{th}(\vec{k}) W + \frac{\delta W}{\delta \vec{\Phi}(\vec{k})} \right]. \tag{A.15}$$

Next, one might expect to conduct a higher-order (i.e., $\vec{k}_2, \vec{k}_3, \cdots$) analysis of the problem in order to determine an equation for $W$. However, even in the discrete case, the kinematic equation (3.9) can be used to reassemble the expansion of $\rho$ and determine a closed equation for $\Psi$ [14-16]. Using (A.8) and (A.15) in (3.9), we find

$$\frac{\partial \Psi}{\partial t} = \int d^3k_1 \omega(\vec{k}_1) \frac{\delta}{\delta \vec{\Phi}(\vec{k}_1)} \int d^{6N}\Gamma^* \Delta(\vec{\Phi} - \vec{\Phi}^*) \vec{\Pi}^*(\vec{k}_1)$$
$$\left( \hat{\rho} W + \int d^3k_2 \omega(\vec{k}_2) \hat{\rho} \int_{-\tau_0}^{0} dt'' e^{-\mathcal{L}_0 t''} \vec{\Pi}(\vec{k}_2) \left[ -\beta \vec{f}^{th}(\vec{k}_2) W + \frac{\delta W}{\delta \vec{\Phi}(\vec{k}_2)} \right] \right) \tag{A.16}$$

and since $\vec{\Pi}^{th} = 0$, this simplifies to

$$\frac{\partial \Psi}{\partial t} = \iint d^3k_1 d^3k_2 \omega(\vec{k}_1) \omega(\vec{k}_2) \frac{\delta}{\delta \vec{\Phi}(\vec{k}_1)} \int d^{6N}\Gamma^* \Delta \vec{\Pi}^*(\vec{k}_1) \hat{\rho} \int_{-\tau_0}^{0} dt'' e^{-\mathcal{L}_0 t''} \vec{\Pi}(\vec{k}_2) \left[ -\beta \vec{f}^{th}(\vec{k}_2) + \frac{\delta}{\delta \vec{\Phi}(\vec{k}_2)} \right] W$$

. (A.17)

Finally, expanding $\Psi$ as we did for $\rho$ in (A.5), it follows that $\Psi \to W$ as $\omega \to 0$ (i.e., when all slow processes operate on a significantly longer timescale than that of atomistic fluctuations), and thus we find

$$\frac{\partial \Psi}{\partial t} = \iint d^3k_1 d^3k_2 \omega(\vec{k}_1) \omega(\vec{k}_2) \frac{\delta}{\delta \vec{\Phi}(\vec{k}_1)} \bullet \left[ \vec{\vec{D}} \left( -\beta \vec{f}^{th}(\vec{k}_2) + \frac{\delta}{\delta \vec{\Phi}(\vec{k}_2)} \right) \Psi \right] \tag{A.18}$$

where $\vec{\vec{D}}(\vec{k}_1, \vec{k}_2) = \int_{-\infty}^{0} dt'' \left[ \vec{\Pi}(\vec{k}_1) e^{-\mathcal{L}_0 t''} \vec{\Pi}(\vec{k}_2) \right]^{th}$. This is the functional Smoluchowski equation for the dynamics of systems with a broad continuum of time scales.

## ACKNOWLEDGEMENTS

This project was supported in part by the US Department of Energy, the National Institutes of Health (NIBIB), the METACyt Initiative, the Office of the Provost at the University of Texas at Arlington, and the College of Arts and Sciences at Indiana University through the Center for Cell and Virus Theory.